\documentclass[journal]{IEEEtran}
\usepackage{url}
\usepackage{amsmath}
\usepackage{graphicx} 
\usepackage{enumitem}
\usepackage{comment}
\usepackage{makecell}
\usepackage{hyperref}
\usepackage{float}
\usepackage[table]{xcolor}
\usepackage{lipsum}
\usepackage{multirow} 
\usepackage{orcidlink}

\begin{document}
\title{Joint Classification and Regression Deep Learning Model for Universal Phase-based Ranging in Multiple Environments}

\author{Pantelis Stefanakis\orcidlink{0009-0008-0401-0229}, and Ming Shen\orcidlink{0000-0002-9388-3513}

\thanks{P. Stefanakis, and M. Shen are with the Department of Electronic Systems, Aalborg University, 9220, Denmark. Email: pantelis.stefanak1@ucalgary.ca.}

}

\maketitle
\begin{abstract}
Phase-Based Ranging (PBR) offers several advantages for estimating distances between wirelessly connected devices, including high accuracy over large distances and the removal of the need for antenna arrays at each transceiver. This study investigates the use of Neural Network (NN)-based models for accurate PBR in three distinct environments: Openfield, Office, and Near Buildings, comparing their performance with established non-NN methods. A novel 2NN Model is proposed, integrating two neural networks: one to classify the environment and another to predict distances. Performance was evaluated over 20 trials for each method and dataset using root mean square error (RMSE) and maximum prediction error.

Results show that the 2NN Model consistently outperformed other methods, frequently ranking among the top methods in minimizing both RMSE and maximum error. In addition, the 2NN Model achieved the best average RMSE and the lowest maximum error. To assess the effect of environment misclassification, filtered versions of the NN models were evaluated by omitting misclassified measurements prior to RMSE calculation. Although unsuitable for production use, the filtered models revealed that misclassifications in the 2NN Model had a significant impact. Its filtered variant achieved the lowest RMSE and maximum error across all datasets, and ranked first in the frequency of attaining the lowest maximum error over 20 trials.

Overall, the findings show that NN models deliver robust, high-accuracy ranging across diverse environments, outperforming non-NN methods and reinforcing their potential as universal PBR solutions when trained on comprehensive distance datasets.
\end{abstract}


\IEEEpeerreviewmaketitle

\section{Introduction} \label{intro}
\IEEEPARstart{T}{he} affordability, energy efficiency, and robust ecosystem of advanced IoT transceivers are pivotal in deploying a range of Real-Time Location Services (RTLS) \cite{6641708}. Advanced IoT transceivers are found in many of our mobile devices to facilitate applications such as secure building access, digital car keys for key-less entry, indoor way-finding, and proximity detection, utilizing precision in localizing devices \cite{thiede2021rtls}. Through the applications listed, IoT transceivers enhance security, navigational capabilities, and personal convenience \cite{bakar2023iot}. 

Among the various ranging techniques utilized by IoT transceivers, Phase-Based Ranging (PBR) has emerged as a superior method, offering highly accurate distance estimations without requiring multiple antennas. This results in simplified hardware design and significant cost savings \cite{zand2019ble}. Traditional techniques like Received Signal Strength Indicator (RSSI), Angle of Arrival (AoA), and Angle of Departure (AoD) have been extensively used for distance estimation \cite{girolami2023bluetooth}. RSSI measures the decay of signal strength over distance but is typically accurate to within several meters, making it less ideal for precise applications \cite{scazzoli2024uav}. AoA and AoD, on the other hand, use triangulation methods with fixed transmitters or receivers equipped with multiple antennas, achieving sub-meter accuracy but at the cost of more complex hardware setups \cite{huang2021performance}.

Unlike AoA and AoD, PBR operates effectively with fewer hardware requirements since it does not require any of its transmitters to be equipped with multiple antennas \cite{zand2019ble, sheikh2023phase}. PBR estimates distances by measuring the time of flight and analyzing the phase of radio signals during PBR events. Each event involves multiple tightly synchronized transmission steps, enabling highly accurate distance measurements, comparable in performance to AoA and AoD \cite{kluge2013patent, channel_sounding}.
This paper covers:
\subsection{Deep Learning Models for PBR}
Recent advancements in machine learning have introduced the possibility of enhancing PBR accuracy through deep learning models \cite{sheikh2023phase}. These models adopted NNs to improve the accuracy of PBR estimations of a non-NN method by filling missing/interfered tones in the frequency spectrum. The NN PBR approaches address the limitations of non-NN PBR methods which often require separate models to yield accurate results in different environments, like having different biases/coefficients for each environment that account for the mutipaths. The NN models dynamically adapt to different settings during training, enhancing both the accuracy and reliability of distance estimations. The proposed 2NN Model integrates two neural networks: one for classifying the environmental context and another for predicting distances based on signal phase information. Machine learning classifiers that utilize signal phase information have been shown to effectively classify environments, successfully distinguishing between measurements with and without multipath components \cite{eriksson2024poster}. In this work, the 2NN Model performance was compared against the 1NN Model which only utilizes a single neural network that performs regression.

\subsection{Comparative Analysis with Conventional Methods}
The 1NN and 2NN models were tested against traditional non-NN methods in diverse environments including an Openfield, an Office, and a Near Building outdoor area. For each method, the overall RMSE value of the predictions for each dataset is calculated, and the maximum prediction errors are identified. These metrics are then used to compare the performance of the methods.

The remainder of the paper is organized as follows: 
In Section \ref{section2}, the system model and PBR are explained, in addition information about the datasets used to conduct the experiments is presented. Section \ref{sectionMethods}, outlines the methods that enable distance estimation after PBR data has been collected. In Section \ref{SectionResults}, the accuracy of distance estimations is compared for all methods presented in section \ref{sectionMethods}. Finally, conclusions are drawn in Section \ref{labelCon}.

\section{Preliminaries} \label{section2}

\subsection{System Model}
When a radio signal propagates through space, it experiences a phase shift, meaning that the measured phase of the signal changes as it travels. This phase shift is influenced by several factors including the distance the signal travels, the propagation medium, and the presence of any objects in the signal's path that might cause reflection, diffraction, or scattering. The phase shift $\Delta\phi$ of a radio wave can be mathematically described as in
\begin{equation}
    \Delta\phi=\frac{2 \pi f D}{c}+\phi_{obj}
    \label{equ5:new456}
\end{equation}
where 
$\Delta \phi$ is the phase shift, $f$ is the frequency of the radio signal, $D$ is the distance the signal has traveled and $\phi_{obj}$ represents any additional phase shift introduced by interactions with objects in the signal's path.

Phase-based ranging involves the exchange of constant tones between two radio transceivers, known as the initiator and reflector, in a two-way communication setup. This process allows for the measurement of the distance between the two devices. By transmitting a continuous tone from the initiator to the reflector and receiving it back, the system evaluates the signal's phase shift to ascertain the distance. Precise measurement of this phase shift, coupled with knowledge of the signal's frequency, allows the system to determine the distance between the two transceivers. The processes of PBR is explained in two steps: 
\subsection*{1) PBR measurement collection}
\begin{figure}[ht]
    \centering
    \includegraphics[width=0.48\textwidth]{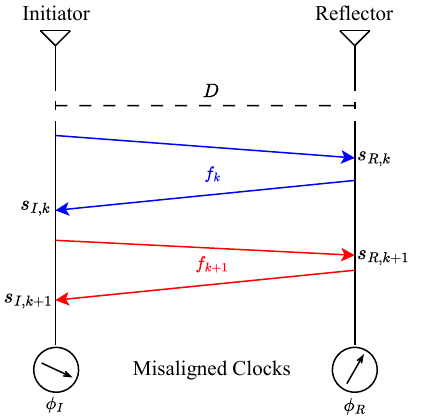}
    \caption{Interaction sequence between the two transceivers.}
    \label{fig:21} 
\end{figure}

The interaction sequence between the two transceivers (the initiator and the reflector) is outlined in Fig. \ref{fig:21}.

Sequence of single measurement procedure:

\begin{enumerate}[label=\arabic*.]
\item The initiator sends out a continuous tone signal $s_k$, on frequency channel $k$.
\item The reflector receives the signal transmitted through the channel $s_{R, k}$, and records the IQ components of the signal together with the known frequency.
\begin{equation}
    \begin{split}
    s_{R,k}=e^{j\left(-\frac{2 \pi f_k D}{c}+\phi_{I,k}-\phi_{R,k}\right)} \\
    \label{equ5:new1}
    \end{split}
\end{equation}
where $D$ denotes the distance between the two transceivers, $f_k$ represents the radio frequency being utilized, $\phi_{I,k}$ signifies the phase of the clock associated with the initiator, while $\phi_{R,k}$ denotes the phase of the clock pertaining to the reflector.

\item The reflector replies to the initiator by sending back a continuous tone signal $s_k$ on the identical frequency channel k.
\item The initiator receives the signal transmitted through the channel $s_{I, k}$, samples and logs the frequency, along with the IQ components of the signal.
\begin{equation}
    \begin{split}
    s_{I,k}=e^{j\left(-\frac{2 \pi f_k D}{c}+\phi_{R,k}-\phi_{I,k}\right)}\\
    \label{equ5:new2}
    \end{split}
\end{equation} 
\item The above sequence is repeated for all available $K$ channels.
\end{enumerate}

Periodically, the data logged in the two transceivers is transferred to a device to process this data and perform phase-based ranging methods. This device can be the initiator or reflector or a third device.

\subsection*{2) PBR measurement processing} 

To simplify \eqref{equ5:new1} and \eqref{equ5:new2}, the expressions are multiplied such that the initiator and reflector clock phases cancel out. This follows the two-way-ranging procedure, in which the initial clock-phase terms introduced by both nodes cancel, leading to the simplified expression:
\begin{equation}
    \begin{split}
    s_{TW,k}=s_{R,k} \cdot s_{I,k}=e^{j\left(-\frac{4 \pi f_k D}{c}\right)}\\
    \label{equ5:new3}
    \end{split}
\end{equation}
Equation \eqref{equ5:new3} follows directly from the two-way-ranging procedure, which is fundamental to the considered system. In a two-way exchange, the initial clock phases of the initiator and the reflector cancel out, yielding the simplified expression. Consequently, the phase of $s_{TW,k}$ becomes
\begin{equation}
    \begin{split}
    \theta_k = \arg(s_{TW,k}) = \left(-\frac{4 \pi f_k D}{c}\right)
    \label{equ5:new4}
    \end{split}
\end{equation}

Equation \eqref{equ5:new4} can be restructured to isolate $D$. However, since $\theta_k$ represents an angle without a defined starting point, it is necessary to calculate two phases to clarify this phase ambiguity: \begin{equation}
    \theta_k = \left(-\frac{4 \pi f_k D}{c}\right)\text{, }
    \theta_{k+1} = \left(-\frac{4 \pi f_{k+1} D}{c}\right)
    \label{equ5:new5}
\end{equation}
Utilizing a pair of points, denoted as $[(f_k, \theta_k), (f_{k+1}, \theta_{k+1})]$, and computing their respective phase as outlined in equation \eqref{equ5:new5} and finding their difference, the resulting equation can be rearrange to isolate $D$. This process results in the formulation
\begin{equation}
    \begin{split}
    \Delta \theta_{k \rightarrow k+1}=\theta_{k+1}-\theta_k&=\left(-\frac{4 \pi D}{c}\right)(f_{k+1}-f_k)\\
    D&=-\frac{c}{4 \pi}\left(\frac{\theta_{k+1}-\theta_k}{f_{k+1}-f_k}\right)\\
    \label{equ5:new6}
    \end{split}
\end{equation}
a linear equation that defines the distance between the two transceivers \cite{zhou2012accurate}.

\subsection{Dataset Collection}
Three datasets were tested to compare the phase-based ranging methods introduced in Section \ref{intro}. These datasets originate from three distinct environments, allowing for an evaluation of each method's robustness in various situations characterized by differing levels of multipath propagation, which can cause fading and phase interference to the signals. Multipath propagation occurs when signals travel from the transmitter to the receiver along different paths due to reflections, diffractions, and scattering from surrounding objects. The environments of the three datasets are:

\begin{itemize}
    \item An Openfield with distances between the two transceivers ranging from 0.1-30.0 m with 0.1 m intervals.
    \item A furnished Office with distances ranging from 0.1-6.2 m with 0.2 m intervals.
    \item An outdoors environment surrounded by buildings with distances ranging from 0.1-10.0 m with 0.2 m intervals.
\end{itemize}

\subsection{Dataset Acquisition} \label{f}
At each distance, multiple measurement procedures are executed and stored with a label denoting the ground-truth distance. Each procedure comprises 72 measurements of the in-phase and quadrature (IQ) components of a tone signal, acquired over 72 distinct channels in the 2.4 GHz ISM band spanning 2.404-2.478 GHz in 1 MHz increments, excluding 2.425, 2.426, and 2.427 GHz. Signals are transmitted according to a measurement-specific frequency-hopping sequence. Consecutive measurement procedures are temporally separated by enforcing a 1-second waiting interval. To collect the measurements, the distance between the two transceivers is adjusted by moving one transceiver towards or away from the other, ensuring that there is always a clear line of sight between them. When the device is being moved the individual holds a button that labels all measurements collected during that time as moving, to discard these measurements.

\section{Methods} \label{sectionMethods}
This section, introduces the methods that enable distance estimation.  
\subsection{Linear Regression}
\begin{figure}[ht]
    \centering
    \includegraphics[width=0.48\textwidth]{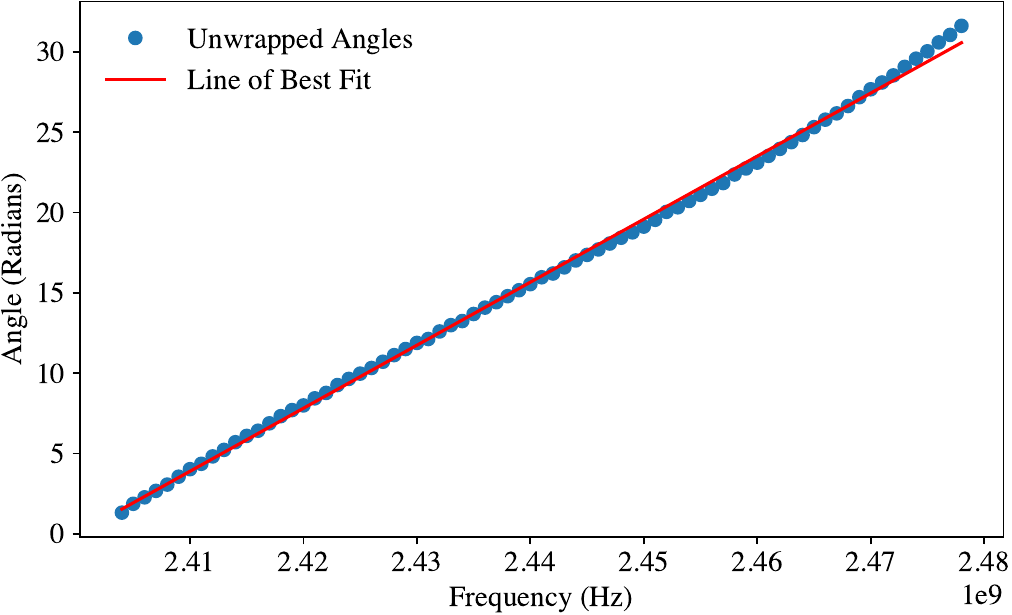}
    \caption{Unwrapped Angles used to estimate the slope from a PBR procedure.}
    \label{fig:R601_2}
\end{figure}

Each PBR procedure contains 72 phase angles, bounded between $-\pi$ and $\pi$, corresponding to a unique frequency, $[(f_0, \theta_0)$, $(f_1, \theta_1)$, $(f_2, \theta_2)$ ...  $(f_{71}, \theta_{71})]$. Fig. \ref{fig:R601_2} shows the angles at each frequency for a single measurement procedure. The blue points represent the 72 distinct frequency channels, while the three red points indicate the interpolated angles for the frequencies 2.425, 2.426, and 2.427 GHz. Although interpolation is shown for visualization purposes, it is not required for phase unwrapping, as the frequency spacing is sufficiently small for the algorithm to accurately resolve phase discontinuities. The unwrapped points form an approximately linear trend, as illustrated in Fig. \ref{fig:R601_2}, consistent with the relationship described in Equation \eqref{equ5:new6}.

\begin{figure}[ht]
    \centering
    \includegraphics[width=0.48\textwidth]{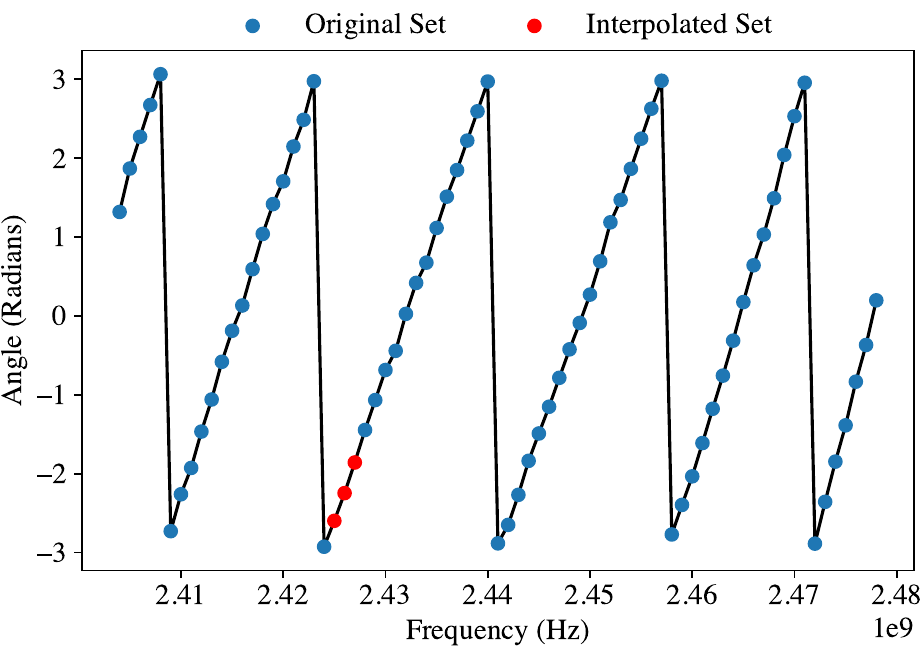}
    \caption{Angles $\theta_k$ from a PBR procedure.}
    \label{fig:R601}
\end{figure}

The distance $D$ is calculated from the slope of the best fit linear line $m$ using equation \eqref{equ5:new61}, where $d$ is the distance corresponding to the delay of the hardware.

\begin{equation}
    D=\frac{c}{4 \pi}|m|-d
    \label{equ5:new61}
\end{equation}

\subsection{Inverse Fast Fourier Transform (IFFT)}
\begin{figure}[ht]
    \centering
        \includegraphics[width=0.48\textwidth]{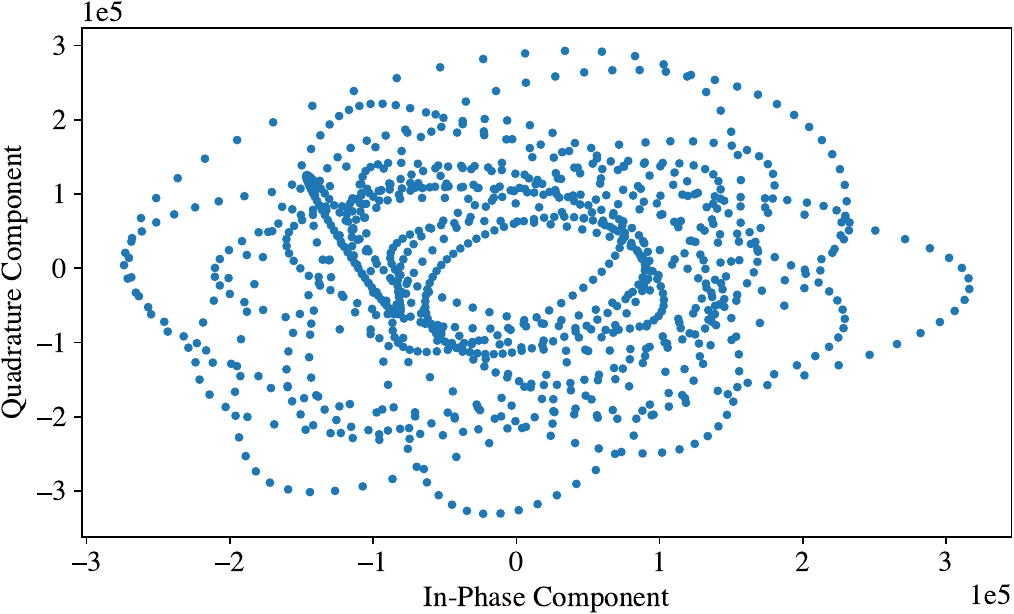}
        \caption{Interpolation model predictions for a specific set of points, applied to a PBR procedure.}
        \label{interpolate}
\end{figure}

\begin{figure}[ht]
        \centering
        \includegraphics[width=0.48\textwidth]{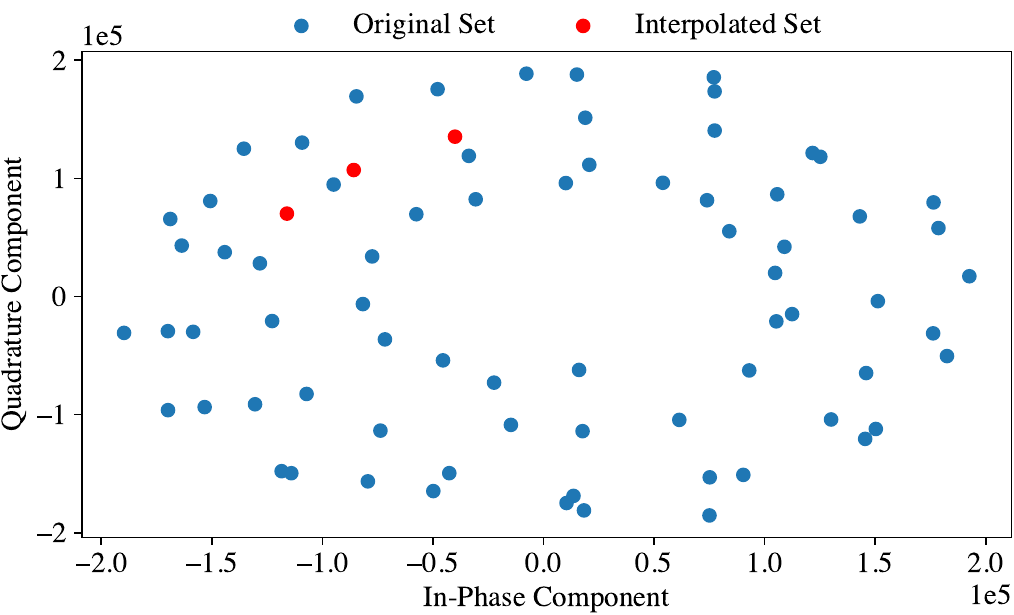}
        \caption{In-phase and quadrature components of initiator/reflector phasors $s_{TW,k}$ across frequency channels. Original measured phasors are shown in blue; interpolated phasors used to replace missing data at 2.425, 2.426, and 2.427 GHz are shown in red.}
    \label{interpolate_2}
\end{figure}

In each Phase-Based Ranging (PBR) procedure, 72 signal representations are captured, each corresponding to a distinct frequency, denoted as {$(f_0, s_{TW,0})$, $(f_1, s_{TW,1})$, ..., $(f_{71}, s_{TW,71})$}. However, due to hardware constraints at three intermediate frequencies 2.425, 2.426, and 2.427 GHz are absent from the dataset. To address the missing values, an interpolation algorithm is employed to estimate the missing signal components. As illustrated in Figs. \ref{interpolate} and \ref{interpolate_2}, the method reconstructs the phasor data at the missing frequencies using surrounding information. This step is critical as the IFFT-based distance estimation technique assumes uniform frequency sampling to accurately transform the frequency-domain signal into the time domain. Missing values violate this assumption, resulting in degraded range resolution or spurious artifacts. Interpolation restores spectral continuity, thereby preserving the fidelity of the reconstructed time-domain signal and ensuring accurate distance estimation.

\begin{figure}[ht]
        \centering
        \includegraphics[width=0.48\textwidth]{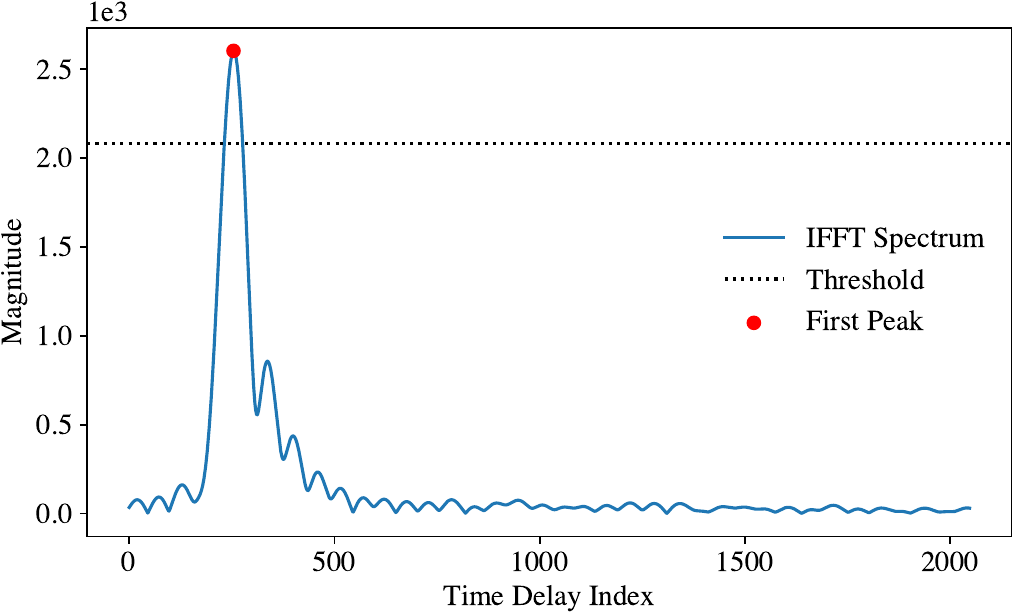}
        \caption{IFFT spectrum used to determine a distance estimate from a PBR procedure following interpolation.}
         \label{fig:R60164_2}
\end{figure}

Following the interpolation, the inverse Fourier transform of the sum of 75 Initiator/Reflector products is computed as:

\begin{equation}
S_{all} = s_{TW,0} + s_{TW,1} + \dots + s_{TW,75}
\end{equation}
Fig. \ref{interpolate_2} illustrates the sum of the 75 Initiator/Reflector products, \(S_{all}\), and Fig. \ref{fig:R60164_2} shows the corresponding IFFT spectrum.

The first peak in the time-domain signal corresponds to the initial arrival of the signal at the transmitter via the line-of-sight path. From the IFFT spectrum, the location of the first peak is selected using a threshold. To relate the position of the first peak in the IFFT spectrum to the time of flight of the signal exchange, it is important to consider the sampling rate and the scale of the time-domain representation. The estimated distance is calculated using the following equations \cite{wehner2011high}:

 \begin{equation}
 \begin{split}
    \Delta'=\frac{1}{TN_{FFT}}=\frac{1}{\left(\frac{1}{f_s}\right) N_{FFT}}
 \end{split}
    \label{fh2}
\end{equation}

 \begin{equation}
     D = \frac{1}{2}\Delta' \tau c-d
    \label{fh3}
 \end{equation}
where $T$ is the sampling period, $\Delta'$ is the frequency resolution, $\tau$ is the time propagation of the signal, $c$ is the speed of light, and $d$ accounts for the delay of the hardware. Increasing \(N_{FFT}\) improves the frequency resolution, allowing for finer distinctions between frequencies in the IFFT output. The frequency resolution, \(\Delta'\), represents the spacing between discrete points in the frequency domain.

\subsection{Neural Network Models}
In this section two NN models are introduced, the 1NN Model, and 2NN Model. Both NN models use a 5\% validation split to continuously evaluate performance on unseen data during training. The model weights corresponding to the lowest validation error on the validation set are selected for final evaluation.

Prior to inputting the data into the models, a structured preprocessing pipeline was employed to transform the raw complex-valued radio signals into informative and standardized feature representations. Each data sample initially consisted of 150 real values, representing 75 complex in-phase (I) and quadrature (Q) signal components. Two complementary feature sets were extracted: IQ features and spectral features. The IQ features were generated by computing the magnitude and phase of the complex signal. Log-magnitude values were calculated to enhance dynamic range stability, while the phase information was encoded using both cosine and sine transformations, resulting in a 225-dimensional feature vector. Concurrently, spectral features were derived by applying an inverse fast Fourier transform (IFFT) to the complex-valued signal, computing the logarithm of the magnitude spectrum, and normalizing the output to zero mean and unit variance, producing a 256-dimensional feature vector. All features and target distances were scaled using standard normalization fitted on the training set alone to avoid information leakage. This preprocessing ensured that the input data captured both time-domain and frequency-domain characteristics.

\subsubsection*{1NN Model}
\begin{figure}[ht]
    \centering
\includegraphics[width=0.48\textwidth]{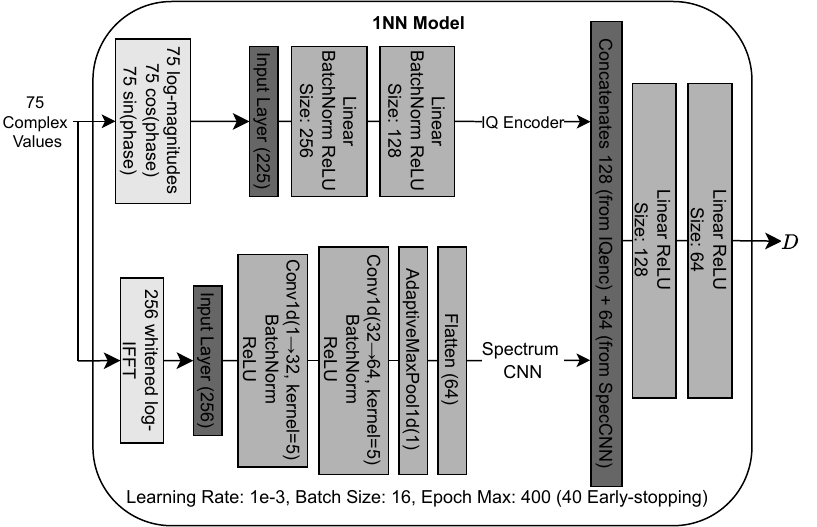}
    \caption{Architecture of the 1NN Model.}
    \label{fig:1NN_Architecture} 
\end{figure}

The 1NN Model takes as input a concatenation of engineered IQ features and spectral features derived from 75 complex in-phase (I) and quadrature (Q) signal components. The network consists of two parallel encoding branches: a fully connected encoder for the 225-dimensional IQ feature vector and a convolutional encoder for the 256-dimensional spectral feature vector. The IQ encoder includes two linear layers with Batch Normalization and ReLU activation, reducing the input to a 128-dimensional representation. The spectral encoder applies two 1D convolutional layers followed by Batch Normalization, ReLU activations, and an adaptive max pooling layer, producing a 64-dimensional feature vector. These embeddings are concatenated and passed through a regression head with three fully connected layers, each followed by ReLU activations except the final output layer. The network is trained using the Smooth L1 loss function, optimized via AdamW, and employs learning rate scheduling and early stopping based on validation RMSE to improve generalization. Fig. \ref{fig:1NN_Architecture}, illustrates the network architecture of the 1NN Model.

\subsubsection*{2NN Model}
The 2NN Model architecture integrates two neural networks: the NN Dataset ID Classifier and the NN With ID Input Model. The first network, the NN Dataset ID Classifier, is trained to classify the environment in which a signal was recorded in, using the same 75 complex-valued IQ signal components as input. These are transformed into engineered IQ and spectral features, then passed through a hybrid neural architecture consisting of a fully connected encoder for IQ features and a convolutional encoder for spectral features. The outputs are concatenated and fed into a classification head with two dense layers, the last using a softmax activation to predict a one-hot environment label. The model is trained using categorical cross-entropy loss and optimized using the Adam optimizer. One-hot encoding of the environment class (e.g.,  is trained on the $s_{TW,k}$ values, with dataset IDs encoded as one-hot vectors: $[0 \ 0 \ 1]$ representing Openfield, $[0 \ 1 \ 0]$ for Office, and $[1 \ 0 \ 0]$ for Near Buildings) ensures class distinction and minimizes classification ambiguity.

\begin{figure}[ht]
    \centering
\includegraphics[width=0.48\textwidth]{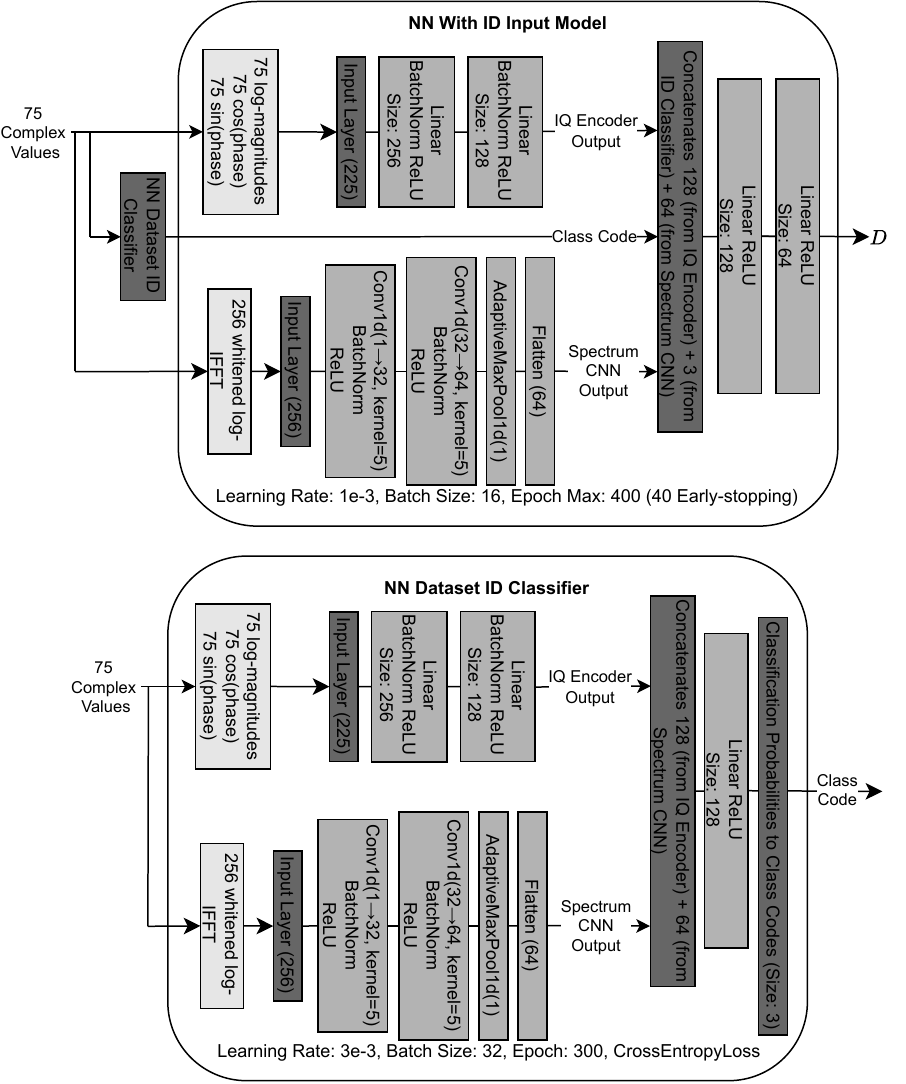}
    \caption{Architecture of the 2NN Model.}
    \label{fig:2NN_IFFT_Architecture} 
\end{figure}

The second network, the NN With ID Input Model, performs distance prediction using the same IQ and spectral features as inputs, but with an important augmentation: the predicted one-hot environment label from the classifier is appended to the feature embedding before regression. This allows the model to leverage environmental context to improve ranging accuracy. The regressor follows a hybrid architecture similar to the 1NN Model, with a fully connected IQ encoder and a spectral convolutional encoder. However, the regression head is modified to accept the combined feature embedding and environment ID vector (192 + 3 = 195 dimensions). The output is a single continuous value representing the predicted distance in millimeters. The model is trained using the Smooth L1 loss function, with early stopping and learning rate scheduling based on validation RMSE to improve generalization. This two-stage architecture allows the 2NN Model to better handle environmental variability by adapting its distance predictions to the predicted signal context. Fig. \ref{fig:2NN_IFFT_Architecture}, illustrates the network architecture of the 2NN Model. 

\section{Results} \label{SectionResults}
\begin{figure}[ht]
    \centering
\includegraphics[width=0.48\textwidth]{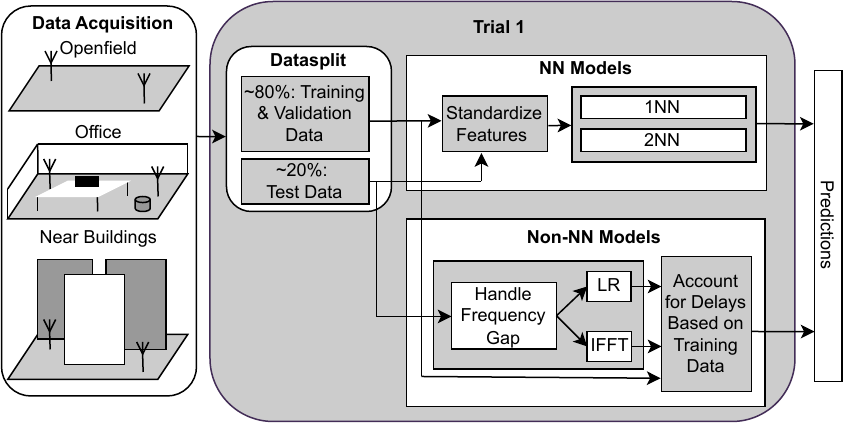}
    \caption{Overview of how distance predictions are made in each trial.}
    \label{fig:OVERVIEW} 
\end{figure}

\begin{figure}[ht]
    \centering
\includegraphics[width=0.48\textwidth]{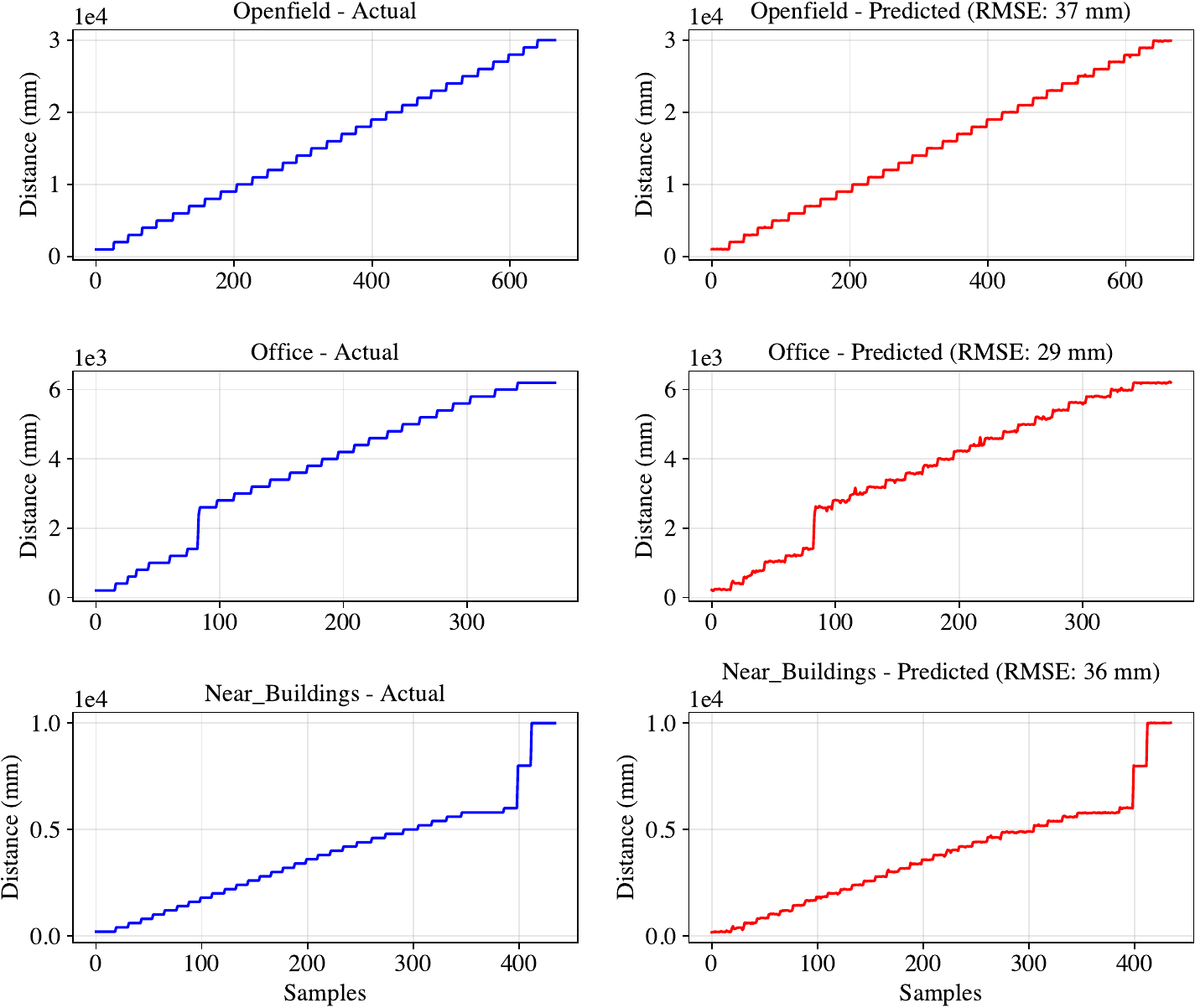}
    \caption{Distance predictions made by the 2NN Model on the three datasets (RMSE Overall: 35 mm).}
    \label{fig:2NN_IFFT_pred} 
\end{figure}

In this section, the performance of ranging using methods presented in Section \ref{sectionMethods} are evaluated. All the methods are trained on around 80\% of the data in their respective dataset. To make sure that all distances in the datasets are included in the training and test data, an algorithm was implemented that randomly assigns around 80\% of the data points at each distance in each of the datasets as training/validation data and the rest as test data. There is variability in distance predictions using NN methods every time the model is trained, attributable to random weight initialization and stochastic training algorithms, and the data splitting. To combat this variability and make an accurate comparison of the methods, the overall RMSE of each PBR method was calculated based on predictions made on 20 trials of random data splitting. Fig. \ref{fig:OVERVIEW}, presents the steps performed in each trial to generate predictions using the methods outlined in section \ref{sectionMethods}. In each trial, random data-splitting was performed, the NN models were trained and all the methods were used to make distance predictions, on the test data. The LR and IFFT methods at each trial utilize the training data to find the bias that accounts for the hardware delay. At the end of each trial, the predictions of each method were saved to enable a comprehensive analysis. Fig. \ref{fig:2NN_IFFT_pred}, shows the distance predictions made by the 2NN Model on the three datasets.

The NN Dataset ID Classifier achieved an accuracy of more than 99\% in classifying inputs in their respective datasets in all 20 trials. Similar to a confusion matrix, Table \ref{tab:class} displays the number of misclassifications for each trial. The varying shades of gray indicate the dataset that each sample was incorrectly classified as.

\definecolor{Openfield}{gray}{0.3}   
\definecolor{Office}{gray}{0.6}      
\definecolor{NB}{gray}{0.85}         

\setlength{\tabcolsep}{3pt}

\begin{table}[ht]
\centering
\caption{}
\label{tab:class}
\begin{tabular}{|c|*{10}{c|}}
\hline
\makecell{\textbf{True} \textbf{Dataset}} & \textbf{1} & \textbf{2} & \textbf{3} & \textbf{4} & \textbf{5} & \textbf{6} & \textbf{7} & \textbf{8} & \textbf{9} & \textbf{10} \\
\hline
\textcolor{Openfield}{Openfield} & 0 & 0 & 0 & 0 & 0 & 0 & 0 & 0 & 0 & 0 \\
\textcolor{Office}{Office} & \textcolor{NB}{1} & 0 & \textcolor{NB}{1} & \textcolor{NB}{1} & 0 & 0 & 0 & 0 & 0 & \textcolor{NB}{1} \\
\textcolor{NB}{\makecell{Near\\Buildings}} & 0 & \textcolor{Office}{1} & 0 & \textcolor{Openfield}{1} & 0 & \textcolor{Office}{1} & \textcolor{Office}{1} & \textcolor{Openfield}{1} & \textcolor{Office}{2} & \textcolor{Office}{1}\\
\hline
\makecell{\textbf{True} \textbf{Dataset}} & \textbf{11} & \textbf{12} & \textbf{13} & \textbf{14} & \textbf{15} & \textbf{16} & \textbf{17} & \textbf{18} & \textbf{19} & \textbf{20} \\
\hline
\textcolor{Openfield}{Openfield} & 0 & 0 & 0 & 0 & 0 & 0 & 0 & 0 & 0 & 0 \\
\textcolor{Office}{Office} & 0 & \textcolor{NB}{1} & \textcolor{Openfield}{1} & \textcolor{NB}{1} & \textcolor{Openfield}{2} & 0 & 0 & 0 & \textcolor{Openfield}{1} & 0 \\
\textcolor{NB}{\makecell{Near\\Buildings}} & \textcolor{Office}{2} & \textcolor{Openfield}{2} & \textcolor{Office}{2} & \textcolor{Office}{1} & \textcolor{Openfield}{1}+\textcolor{Office}{1} & \textcolor{Office}{1} & \textcolor{Openfield}{1} & \textcolor{Openfield}{1}+\textcolor{Office}{1} & \textcolor{Office}{1} & \textcolor{Office}{1} \\
\hline
\end{tabular}
\end{table}

Table \ref{tab:ranking} presents the overall RMSE for each method and dataset. The RMSE is computed by first concatenating the predictions for each method in the different datasets, calculating the RMSE on the combined data for each of the 20 trials, and then averaging the results across trials. The label "Filtered" after the 1NN and 2NN indicates that the misclassified measurements by the NN Dataset ID Classifier were omitted before the RMSE was calculated. This was done to investigate the impact of misclassifications on distance estimation. However, these methods are not suitable for production use, as there is no way to determine in advance whether a measurement will be correctly classified. The labels "All", "Openfield", "Office", and "NB" following LR and IFFT indicate the training dataset(s) used to determine the bias. In the same table, it is evident that the 2NN method attained the lowest overall RMSE in all datasets excluding the Office in which the 1NN method beat it by 1 mm. Table \ref{tab:ranking}, also shows that the NN methods are leading the non-NN methods in making predictions with lower RMSEs in all datasets. In addition, misclassified measurements showed to have a significant impact in the RMSE ranging prediction performance of the 2NN method as the 2NN Filtered method achieved significantly lower overall RMSEs than all the other methods in all datasets.
\begin{table}[ht] 
\centering
\caption{}
\begin{tabular}{|c|c|c|c|c|}
\hline
\cellcolor{black}{} & 
\makecell{\textbf{Openfield} \\ Overall \\ RMSE \\ (mm)} & \makecell{\textbf{Office} \\ Overall \\ RMSE \\ (mm)} & \makecell{\textbf{Near}\\ \textbf{Buildings} \\ Overall \\ RMSE \\ (mm)}  & \makecell{\textbf{Combined} \\ Overall \\ RMSE \\ (mm)} \\ \hline 
\makecell{\textbf{1NN}}  & \makecell{54} & \makecell{47} & \makecell{56} & \makecell{56} \\ \hline
\makecell{\textbf{2NN}} & \makecell{38} & \makecell{48} & \makecell{54} & \makecell{49} \\ \hline
\makecell{\textbf{1NN Filtered}}  & \makecell{54} & \makecell{46} & \makecell{49} & \makecell{54} \\ \hline
\makecell{\textbf{2NN Filtered}} & \makecell{38} & \makecell{32} & \makecell{43} & \makecell{40} \\ \hline
\makecell{\textbf{LR All}} & \makecell{224} & \makecell{750} & \makecell{435} & \makecell{473} \\ \hline
\makecell{\textbf{LR Openfield}} & \makecell{167} & \makecell{843} & \makecell{382} & \makecell{490} \\ \hline
\makecell{\textbf{LR Office}} & \makecell{666} & \makecell{556} & \makecell{821} & \makecell{691} \\ \hline
\makecell{\textbf{LR NB}} & \makecell{204} & \makecell{921} & \makecell{369} & \makecell{526} \\ \hline
\makecell{\textbf{IFFT All}} & \makecell{214} & \makecell{948} & \makecell{391} & \makecell{541} \\ \hline
\makecell{\textbf{IFFT Openfield}} & \makecell{45} & \makecell{1107} & \makecell{305} & \makecell{581} \\ \hline
\makecell{\textbf{IFFT Office}} & \makecell{886} & \makecell{645} & \makecell{939} & \makecell{849} \\ \hline
\makecell{\textbf{IFFT NB}} & \makecell{66} & \makecell{1147} & \makecell{301} & \makecell{601} \\ \hline
\end{tabular}
\label{tab:ranking}
\end{table}

\begin{figure}[ht]
    \centering
\includegraphics[width=0.48\textwidth]{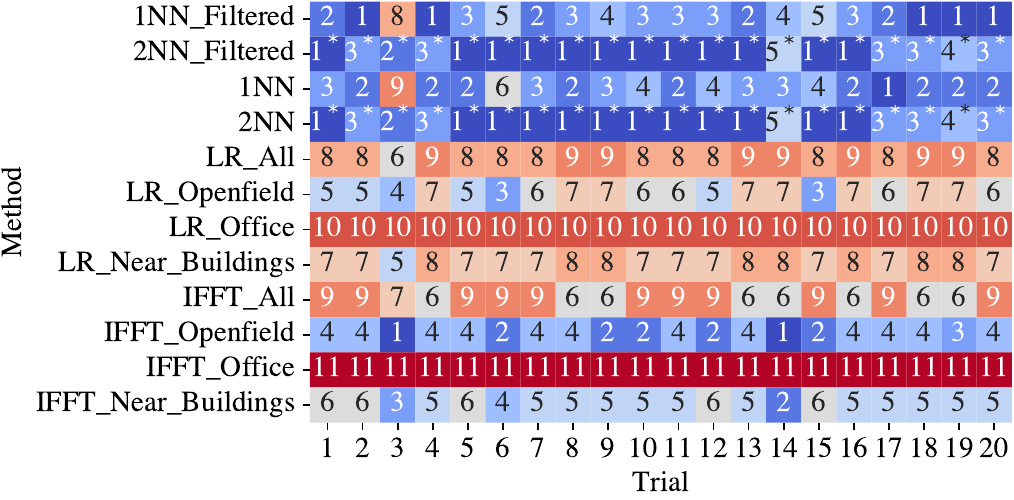}
    \caption{RMSE Ranking from best to worst 1-12 performing PBR methods in the Openfield environment.}
    \label{fig:Ranking_Op} 
\end{figure}

\begin{figure}[ht]
    \centering
\includegraphics[width=0.48\textwidth]{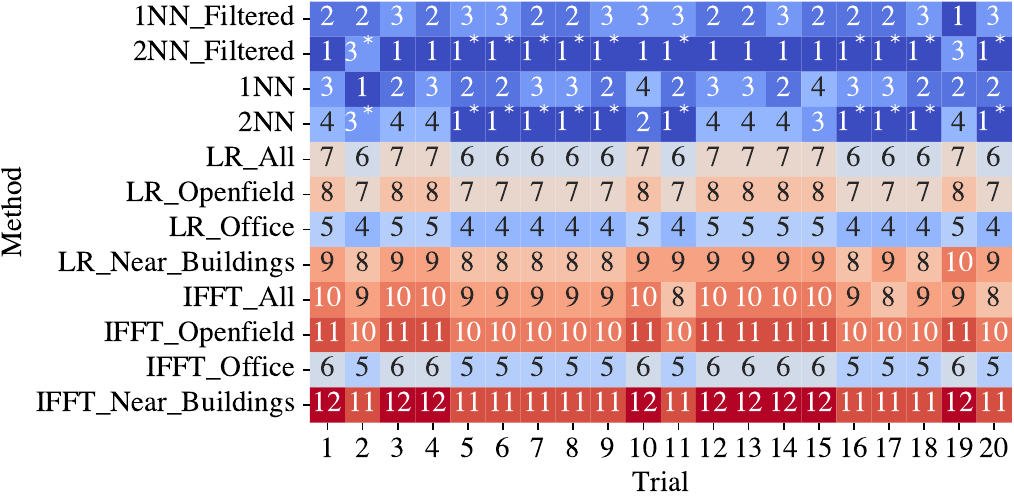}
    \caption{RMSE Ranking from best to worst 1-12 performing PBR methods in the Office environment.}
    \label{fig:Ranking_Of} 
\end{figure}

\begin{figure}[ht]
    \centering
\includegraphics[width=0.48\textwidth]{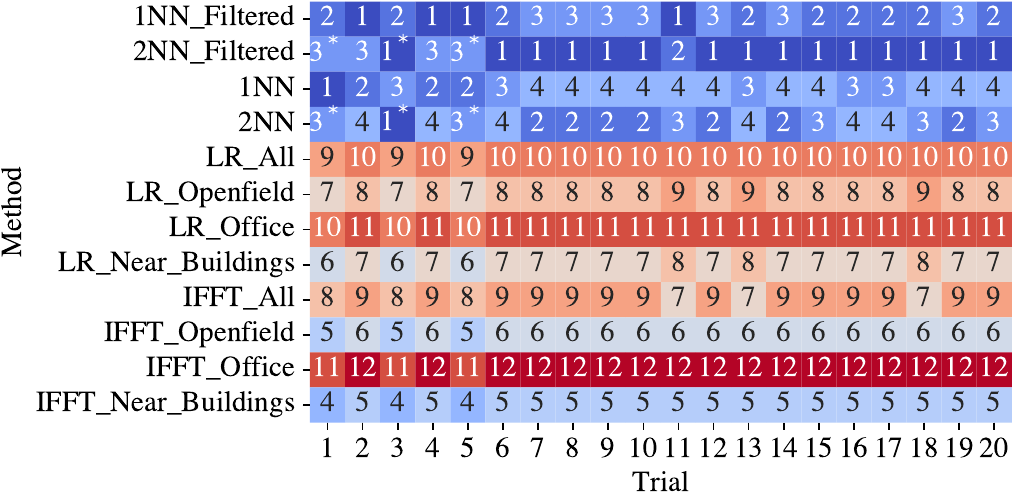}
    \caption{RMSE Ranking from best to worst 1-12 performing PBR methods in the Near Buildings environment.}
    \label{fig:Ranking_Nb} 
\end{figure}

The ranking based on RMSE performance of the distance predictions made by the PBR methods in each of the datasets-Openfield, Office, and Near Buildings in 20 trials can be visualized in Figs. \ref{fig:Ranking_Op}, \ref{fig:Ranking_Of} and \ref{fig:Ranking_Nb}. Across all datasets, excluding the filtered methods, the 2NN method most frequently achieved the lowest RMSE. Specifically, in the Openfield dataset, 2NN led in 12 out of 20 trials, followed by 1NN in 6 and IFFT in 2. In the Office dataset, 2NN outperformed in 12 trials, while 1NN led in the remaining 8. For the Near Buildings dataset, 2NN was best in 12 trials, with 1NN taking 8. In the combined dataset, 2NN again dominated with 13 lowest-RMSE results, compared to 1NN's 7. Notably, except for Openfield, the competition for best performance was consistently between 1NN and 2NN, which always ranked first or second in every trial when filtered methods were excluded.

\begin{figure}[ht]
    \centering
    \includegraphics[width=0.48\textwidth]{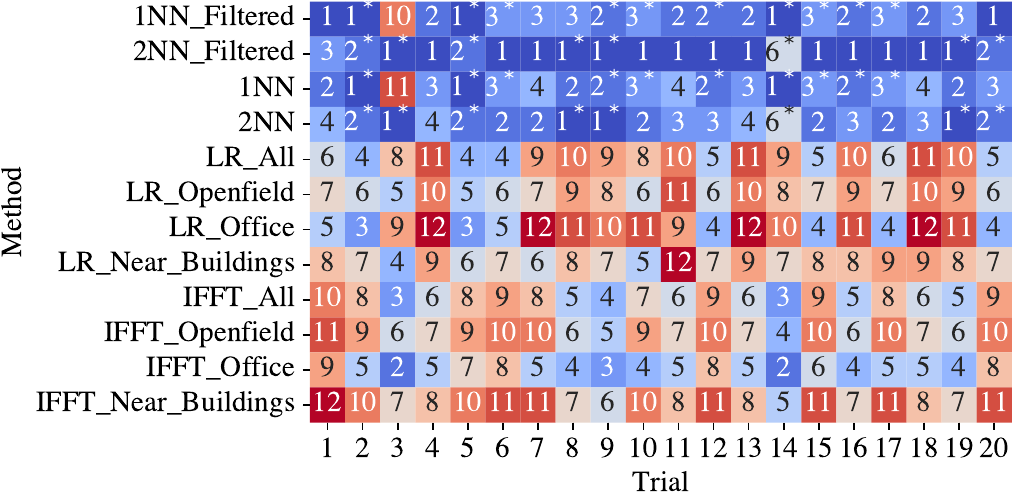}
    \caption{Lowest Maximum Error Ranking from best to worst 1-12 performing PBR methods in the combined dataset.}
    \label{fig:max_error} 
\end{figure}

Table \ref{tab:ranking_Max} ranks the methods by the frequency with which they achieved the lowest maximum distance prediction error across 20 trials. When the filtered methods are excluded, the 2NN Model most often achieved the lowest maximum error on average, followed by the 1NN Model. In addition, the 2NN Model recorded a lower overall maximum error than all other methods. Figure \ref{fig:max_error} visualizes the rankings for each method and trial based on maximum error. In the combined dataset (excluding filtered methods), the 2NN Model achieved the lowest maximum error in 12 out of 20 trials, while the 1NN Model did so in the remaining 8 trials. Across all trials, the 1NN and 2NN models consistently ranked first or second in terms of lowest maximum error.
Moreover, misclassified measurements were found to have a significant impact on the maximum error in distance prediction performance of the 2NN method, as the 2NN Filtered method achieved a significantly lower maximum error than all the other methods and ranked first in the frequency with which it achieved the lowest maximum distance prediction error across the 20 trials.

\begin{table}[ht]
\centering
\caption{}
\label{tab:avg_ranking}
\begin{tabular}{|c|c|c|}
\hline
\cellcolor{black}{} & \makecell{ \textbf{Average Rank (Smallest Max Error)}}  & \makecell{\textbf{Max Error}} \\ \hline
\textbf{2NN Filtered} & 1.50 & 627 mm \\ \hline
\textbf{1NN Filtered} & 2.50 & 1113 mm \\ \hline
\textbf{2NN} & 2.50 & 1040 mm \\ \hline
\textbf{1NN} & 2.95 & 1227 mm\\ \hline
\textbf{IFFT Office} & 5.20 & 2784 mm \\ \hline
\textbf{IFFT All} & 6.70 & 3459 mm\\ \hline
\textbf{LR NB} & 7.55 & 4655 mm\\ \hline
\textbf{LR Openfield} & 7.60 & 4667 mm\\ \hline
\textbf{LR All} & 7.75 & 4679 mm\\ \hline
\textbf{IFFT Openfield} & 7.95 & 3668 mm\\ \hline
\textbf{LR Office} & 8.10 & 4793 mm\\ \hline
\textbf{IFFT NB} & 8.95 & 3718 mm \\ \hline
\end{tabular}
\label{tab:ranking_Max}
\end{table}

Collecting measured data across all possible environments and distances is time-consuming. Therefore, it is valuable to investigate how neural network methods perform when trained on data that excludes the distances present in the test set. To test this situation, the NN models presented in section \ref{sectionMethods} were trained on half of the distances in each dataset and tested on the rest of them. The data was split in an alternating pattern based on the actual distance measured: for instance, the smallest distance was assigned to the training set, the second smallest to the test set, and this pattern continued throughout the dataset.

Table \ref{tab:table2} shows that training on only half of the distance configurations leads to high RMSE on the unseen configurations, indicating poor prediction accuracy. Both NN models, perform significantly worst than the models trained on all distance setups. The NN Dataset ID Classifier in the 2NN Model is not able to classify between the environments with an accuracy below 50\%. The poor performance in distance predictions indicates that even in the same environment, the multi-paths of the signals vary significantly between small distance intervals (1 cm). 

\begin{table}[ht] 
\centering
\caption{}
\begin{tabular}{|c|c|c|c|c|}
\hline
\cellcolor{black}{} & \makecell{\textbf{Openfield}} & \makecell{\textbf{Office}} & \makecell{\textbf{Near Buildings}} & \makecell{\textbf{Average}}  \\ \hline 
\begin{tabular}{@{}c@{}} 1NN Model \end{tabular} & 14586 mm &639 mm& 1013 mm& 10064 mm\\ \hline
\begin{tabular}{@{}c@{}} 2NN Model \end{tabular} &   \multicolumn{4}{c|}{\makecell{Classifier Failed (\textless 50\% accuracy)}} \\ \hline
\end{tabular}
\label{tab:table2}
\end{table}

\section{Conclusion} \label{labelCon}
This paper presents a comprehensive study on the performance of various PBR techniques, with its primary contribution being the introduction of a novel NN-based PBR method that employs two neural networks: one to classify the environment and another to predict range values through regression. The NN Dataset ID Classifier, used to distinguish between three different environments, achieved an accuracy of at least 99\% in all 20 trials.

From the results collected over these trials, the 2NN Model achieved the best overall results among both NN and non-NN methods, with an average RMSE of 49 mm, followed by the 1NN Model at 56 mm. However, the 1NN Model performed slightly better than the 2NN Model in the Office environment. The weaker performance of the 2NN Model in some trials can be attributed to environment misclassifications, which were found to have a significant impact on prediction accuracy.

Evaluation across individual and combined datasets consistently highlights the strong performance of both 1NN and 2NN. In terms of RMSE, the 2NN Model frequently achieved the lowest values, with the 1NN Model often following closely. When considering maximum prediction error, the 2NN Model outperformed the other methods on the combined dataset and ranked slightly higher in the frequency with which it achieved the lowest maximum distance prediction error across the 20 trials. Notably, in nearly all trials, the 1NN and 2NN methods consistently ranked at the top, demonstrating robustness and reliability.

The results suggest that 1NN and 2NN provide a universal solution for ranging estimation across diverse environments, offering superior prediction performance compared to non-NN methods. Finally, it was concluded that NN models can accurately predict distances between two transceivers without antenna arrays from PBR data, provided that all distance setups are represented in the model's training dataset.

\section*{Acknowledgment}
The authors would like to express their sincere gratitude to Prof. Petar Popovski, Department of Electronic Systems, Aalborg University, for his invaluable guidance, encouragement, and mentorship during the development and completion of this journal article.

\ifCLASSOPTIONcaptionsoff
  \newpage
\fi

\end{document}